\documentclass{INTERSPEECH2023}

 \interspeechcameraready

\usepackage[utf8]{inputenc}
\usepackage{algorithm}
\usepackage{algorithmic}
\usepackage{cite}
\usepackage{bm}
\usepackage{hyperref}
\usepackage{comment}
\usepackage{subcaption}

\title{Retraining-free Customized ASR for Enharmonic Words Based on a Named-Entity-Aware Model and Phoneme Similarity Estimation}

\name{Yui Sudo$^1$, Kazuya Hata$^1$, Kazuhiro Nakadai$^{2}$}
\address{
  $^1$Honda Research Institute Japan Co., Ltd., Saitama, Japan\\
  $^2$Department of Systems and Control Engineering, School of Engineering, \\Tokyo Institute of Technology, Tokyo, Japan}
\email{\{yui.sudo, kazuya.hata\}@jp.honda-ri.com, nakadai@ra.sc.e.titech.ac.jp}

\begin{document}

\maketitle
\begin{abstract}
End-to-end automatic speech recognition (E2E-ASR) has the potential to improve performance, but a specific issue that needs to be addressed is the difficulty it has in handling enharmonic words: named entities (NEs) with the same pronunciation and part of speech that are spelled differently. This often occurs with Japanese personal names that have the same pronunciation but different Kanji characters. Since such NE words tend to be important keywords, ASR easily loses user trust if it misrecognizes them. To solve these problems, this paper proposes a novel retraining-free customized method for E2E-ASRs based on a named-entity-aware E2E-ASR model and phoneme similarity estimation. Experimental results show that the proposed method improves the target NE character error rate by 35.7\% on average relative to the conventional E2E-ASR model when selecting personal names as a target NE. 
\end{abstract}


\section{Introduction}
\label{sec: intro}

End-to-end automatic speech recognition (E2E-ASR) has been intensively studied~\cite{ctc1,
rnnt1,ctc2,attention2,attention1,watanabe2017hybrid}. 
This approach combines an acoustic model (AM) and a language model (LM) into a single neural network-based model to improve ASR performance.
However, in practical applications of the E2E-ASR system, a specific challenge arises regarding low customizability. 
As such a problem, we focus on handling words with the same pronunciation and part of speech but are spelled differently, which we call ``enharmonic words.''
The enharmonic words commonly appear in Japanese person names that have the same pronunciation but are represented using different Kanji characters, such as ``Abe'' (a Japanese person name), which can be multiply represented as follows:

\vspace*{-4mm}
\begin{figure}[h]
\begin{center}
{
\resizebox*{5.5cm}{!}{\includegraphics{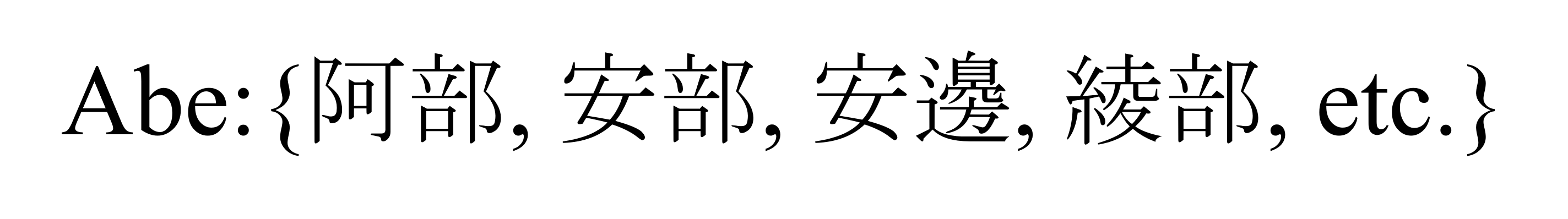}}}
\label{enharmonic} 
\end{center}
\vspace*{-9mm}
\end{figure}

\noindent 
Personal names are an important named entity (NE), so mis-recognition of such words offends users and reduces their trust in the ASR system. In addition, there may be multiple persons in the same group or section with names that read the same but have different Kanji characters. In such cases, the mis-recognition of the names can confuse people.
These problems are not limited to Japanese, but also exist in other languages more or less in writing of personal names.
Since two enharmonic words have the same part of speech ({\it e.g.} person's name), they are more difficult to handle compared to two homonyms, which may have different parts of speech and can be solved in context.
Moreover, because enharmonic words may not be included in the training data, in which case this enharmonic words problem typically includes both in-vocabulary (IV) and out-of-vocabulary (OOV) NE words.

\begin{figure}
\vspace*{-2mm}
\begin{center}
{
\resizebox*{7.5cm}{!}{\includegraphics{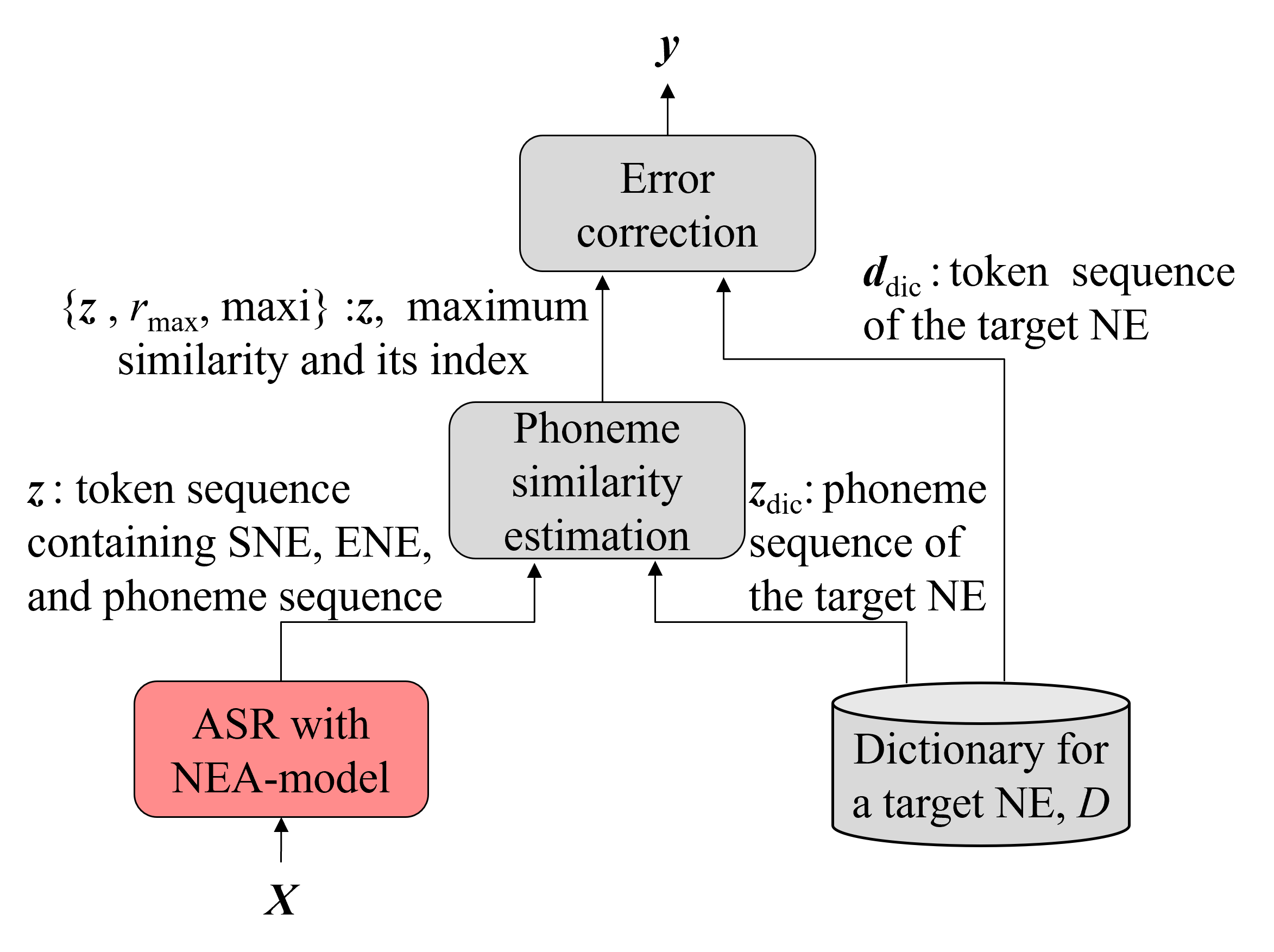}}}
\vspace*{-4mm}
\caption{Overall architecture of the proposed method which consists of ASR with a NEA ASR model, a dictionary, a phoneme similarity estimation, and an error correction.}
\label{overall} 
\end{center}
\vspace*{-9mm}
\end{figure}

A number of  methods based on an weighted finite state transducer (WFST) have been proposed to address the customizability issues for non-E2E-ASR \cite{bazzi00_icslp,Yazgan2004HybridLM,9053872,Brown1992ClassBasedNM,Horndasch2016HowTA}.
The WFST has been used to address pure OOV problems, but it is difficult to address enharmonic word problems \cite{bazzi00_icslp,Yazgan2004HybridLM,9053872}. The class N-gram \cite{Brown1992ClassBasedNM,Horndasch2016HowTA} allows users to register NE words without huge computational cost such as retraining neural networks,  which makes it easier to handle enharmonic words whether they are IV or OOV, because the phoneme sequences can be obtained separately from the AM. However, E2E-ASR is preferable due to its high performance. Methods combining E2E-ASR with WFST have been reported. For example, \cite{Huang2020ClassLA,47384,kojima2022} used DNN-WFST models to detect NE words and construct in-class LMs for NE words. Although these methods enable the customization of the target NE words using in-class LMs, it is unable to handle enharmonic words. This is because both phonemes and tokens are necessary to deal with such words while they output only a sequence of tokens rather than that of phonemes.

Several E2E-ASR-based methods without WFST for the customization have been proposed. These can be divided into two categories: retraining-based or retraining-free. 
The first E2E-based approach involves retraining the E2E-ASR model using text data or an LM. One such method is LM fusion~\cite{Kannan2018AnAO,sriram2018cold}. 
This method attempts to improve accuracy by combining the E2E-ASR model with another LM and by rescoring the N-best hypotheses generated by the E2E-ASR.
However, the effect of LMs connected to the AM in a cascade way is limited, because the original E2E-ASR model is not retrained, and thus the hypotheses to be generated are not updated.
Domain adaptation, which has been investigated to overcome this problem, involves directly retraining the ASR model using only text data ~\cite{Pylkknen2021FastTD,ChenMPL22,sato22_interspeech}. 
Compared to LM fusion, errors are less likely to occur, since the E2E-ASR model is directly updated, while the retraining is time-consuming. In addition to these two methods, ASR model adaptation using speech synthesis (TTS) has been reported~\cite{Sim2019PersonalizationOE,Zheng2021,deng2021improving}. This TTS-based adaptation uses only text data, but it requires a TTS model. Although these retraining-based methods can achieve the customization, they require certain amount of retraining data and considerable retraining time.

The second approach for the E2E-based customization is retraining-free. 
Knowledge-based modeling~\cite{Das2022} and tree-based contextualization~\cite{sun2021tree} efficiently represent relationships between words through knowledge graphs and tree structures, enabling rare word recognition.
Biasing~\cite{deepcontext2018,Jain2020ContextualRF,bruguier2019phoebe} allows users to register any phrases in an editable phrase list to bias the ASR model towards the registered phrases.
Another approach is the contextual adapter~\cite{Sathyendra2022}, which enables users to register arbitrary words in a catalog list without retraining using a training adapter coupled to a pre-trained E2E-ASR with a small amount of speech data.
Although these methods offer flexible customization, they cannot handle pairs of the token and phoneme sequences, which is necessary to handle enharmonic words.

Therefore, this paper proposes a novel customized E2E-ASR model that can handle enharmonic words without the need for retraining. This method is designed to be accessible to users without expertise in ASR or linguistics. The proposed method leverages a named-entity-aware (NEA) ASR model to extract target named entities (NEs) or proper nouns. It then estimates the phoneme similarity between the extracted NE and each word entry in a dictionary containing enharmonic words. If the similarity is high enough, the extracted phoneme sequence is replaced with the dictionary word that has the highest similarity, allowing for improved recognition of the target NE.

\section{Preliminary}
\label{sec:Preliminary}

This section briefly introduces the attention-based Conformer~\cite{gulati2020conformer,guo2021recent}, which is used with the proposed method in the ASR system.
The Conformer encoder consists of two convolutional layers, a linear projection layer, and a positional encoding layer, followed by Conformer blocks. 
The Conformer blocks transform an audio feature sequence, \begin{math}\bm{X}\end{math}, into a hidden state 
\begin{math}
\bm{H}
\end{math} as,
\begin{align}
\label{transformer-encoder}
\bm{H} & = \mathrm{Encoder}(\bm{X}).
\end{align}
Each Conformer block has a multiheaded self-attention layer, a convolution layer, two linear layers, and a layer-normalization layer~\cite{ba2016layer}, with residual connections~\cite{7780459}. 
Given \begin{math}\bm{H}\end{math} generated by the encoder in Eq.~\eqref{transformer-encoder} and a previously estimated token sequence \begin{math}\bm{y}_{s-1}=\{y_1,\cdots,y_{s-1}\}\end{math}, the decoder estimates the next token \begin{math}y_s\end{math} where \begin{math}s\end{math} represents an output label index. This process is recursively performed as follows:
\begin{align}
y_s & = \mathrm{Decoder}(\bm{y}_{s-1}, \bm{H}).
\end{align}
The previously estimated token sequence $\bm{y}_{s-1}$ is first converted into token embeddings. These are then fed into decoder layers with hidden states $\bm{H}$, followed by a linear projection. The predicted probability of $y_s$ is obtained using a softmax function, given outputs of the linear projection. The decoder layer comprises a self-attention network and source-target attention, followed by a position-wise feed-forward network. 
The likelihood of the total token sequence $\bm{y}_S$ consisting of $S$ tokens estimated by an attention model is described as follows:
\begin{equation}
\label{attlikelihood}
P(\bm{y}_{\mathrm{S}} \mid \bm{X})=\prod_{s=1}^{S} P\left(y_{s} \mid \bm{y}_{s-1}, \bm{H} \right).
\end{equation}

\section{Proposed method}
\label{sec: proposed method}
The overall architecture of the proposed method is shown in Figure \ref{overall}. It consists of ASR with a NEA model trained on a corpus tagged with target NEs or proper nouns: a dictionary for target NEs containing enharmonic words, phoneme similarity estimation, and error correction.
From the input signal $\bm{X}$, instead of $\bm{y}_S$, the ASR with NEA model first 
estimates an NE-aware token sequence $\bm{z}$ defined as follows:
\begin{eqnarray}\nonumber
    \bm{z} & = & \{y_1, \cdots, y_{s_1-1}, \{\text{`}<\text{'}, 
    y_{s_1}, \cdots, y_{e_1}, \bm{z}_{\text{p}_1}, \text{`}>\text{'}\},\\\nonumber
    && y_{e_1+1}, \cdots, y_{s_n-1}, \{\text{`}<\text{'}, y_{s_n}, \cdots, y_{e_n}, \bm{z}_{\text{p}_n}, \text{`}>\text{'}\},\\
    && y_{e_n+1}, \cdots, 
    y_S\}.
\end{eqnarray}%
where `$<$' and `$>$' represent special tokens that denote the start of NE (SNE) and end of NE (ENE), and
$y_{s_n}$ and $y_{e_n}$ $(n = 1,\cdots,N)$ are the start and end tokens for the $n$-th NE token sequence $\bm{y}_{\text{NE}_n}$ in $\bm{z}$. 
$\bm{z}_{\text{p}_n}$ is the phoneme sequence corresponding to $\bm{y}_{\text{NE}_n}$. 
For example, given the ground truth ``my name is $y_1 y_2 y_3$,'' we simply insert a special token representing the start and end of the NE word and the corresponding phoneme sequence, as in "my name is $<y_1 y_2 y_3$, $z_1 z_2 z_3>$." Since the part-of-speech and pronunciation are provided by the dataset, the NEA model can be easily trained.
The phoneme similarity estimation then calculates the similarity in the phoneme sequence between the extracted phoneme sequence $\bm{z}_{\text{p}_n}$ and each word entry in the dictionary, $D$, defined as follows: 
\begin{eqnarray}
\label{eq:dictionary}
    D &=& \{\{\bm{d}_{\text{NE}_i},\bm{z}_{\text{d}_i}\}|i=1,\cdots,I\}, 
\end{eqnarray}%
where $\bm{d}_{\text{NE}_i}$ and $\bm{z}_{\text{d}_i}$ represent the $i$-th NE token sequence and phoneme sequence in $D$, respectively.

Since the method comprises simple pairs of NE tokens and phoneme sequences, users without expertise in linguistics can easily edit entries. 
When considering that $\text{maxi}$ is the index of $D$ with the highest similarity,
the error correction replaces the extracted phoneme sequence, $\bm{z}_{\text{p}_n}$, with the dictionary word with the highest similarity $\bm{d}_{\text{NE}_\text{maxi}}$ to correct for ASR errors caused by enharmonic words, when the similarity is higher than the threshold, $V_{th}$.
To correctly estimate the phoneme sequences of NEs, we propose two network models: single-encoder NEA (S-NEA) and dual-encoder NEA (D-NEA).

\begin{figure}[t!]
     \centering
     \hfill
     \begin{subfigure}[b]{0.49\linewidth}
         \centering
         \includegraphics[scale=0.29]{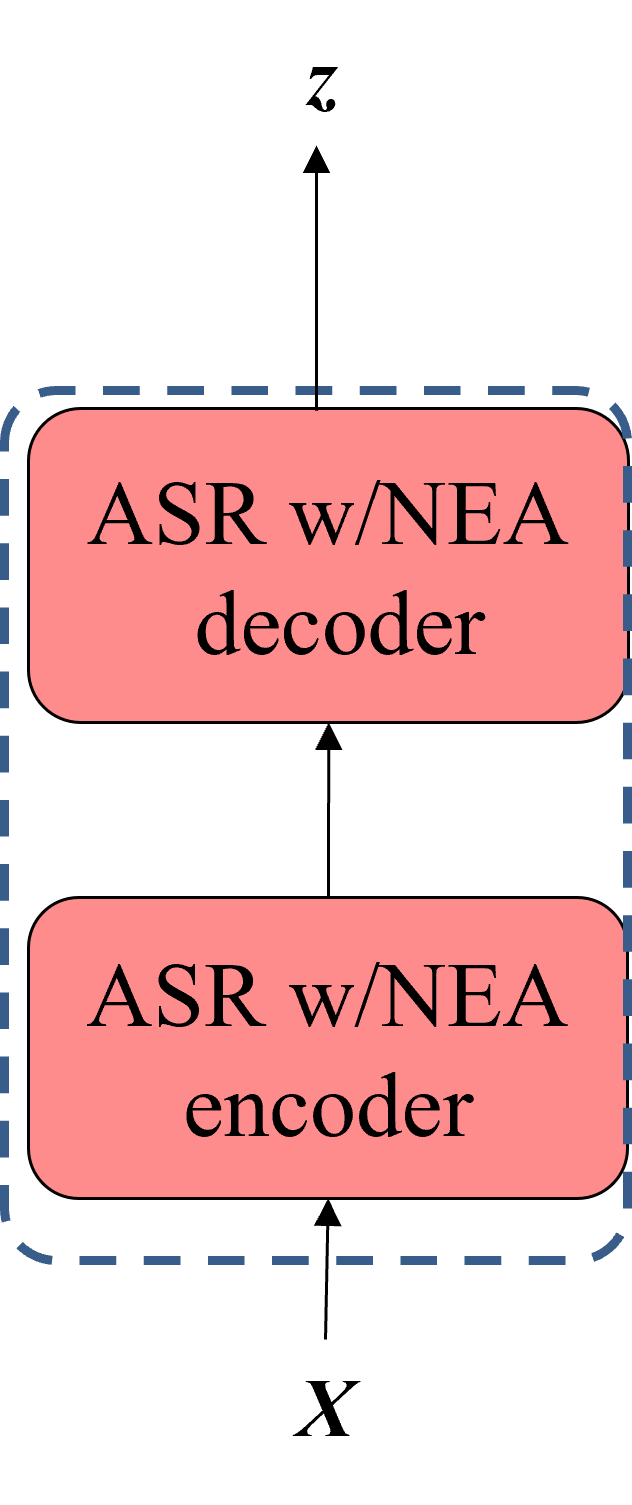}
         \vskip -0.1in
         \caption{Single-encoder NEA.}
         \label{fig:snea}
     \end{subfigure}
     \hfill
     \begin{subfigure}[b]{0.49\linewidth}
         \centering
         \includegraphics[scale=0.29]{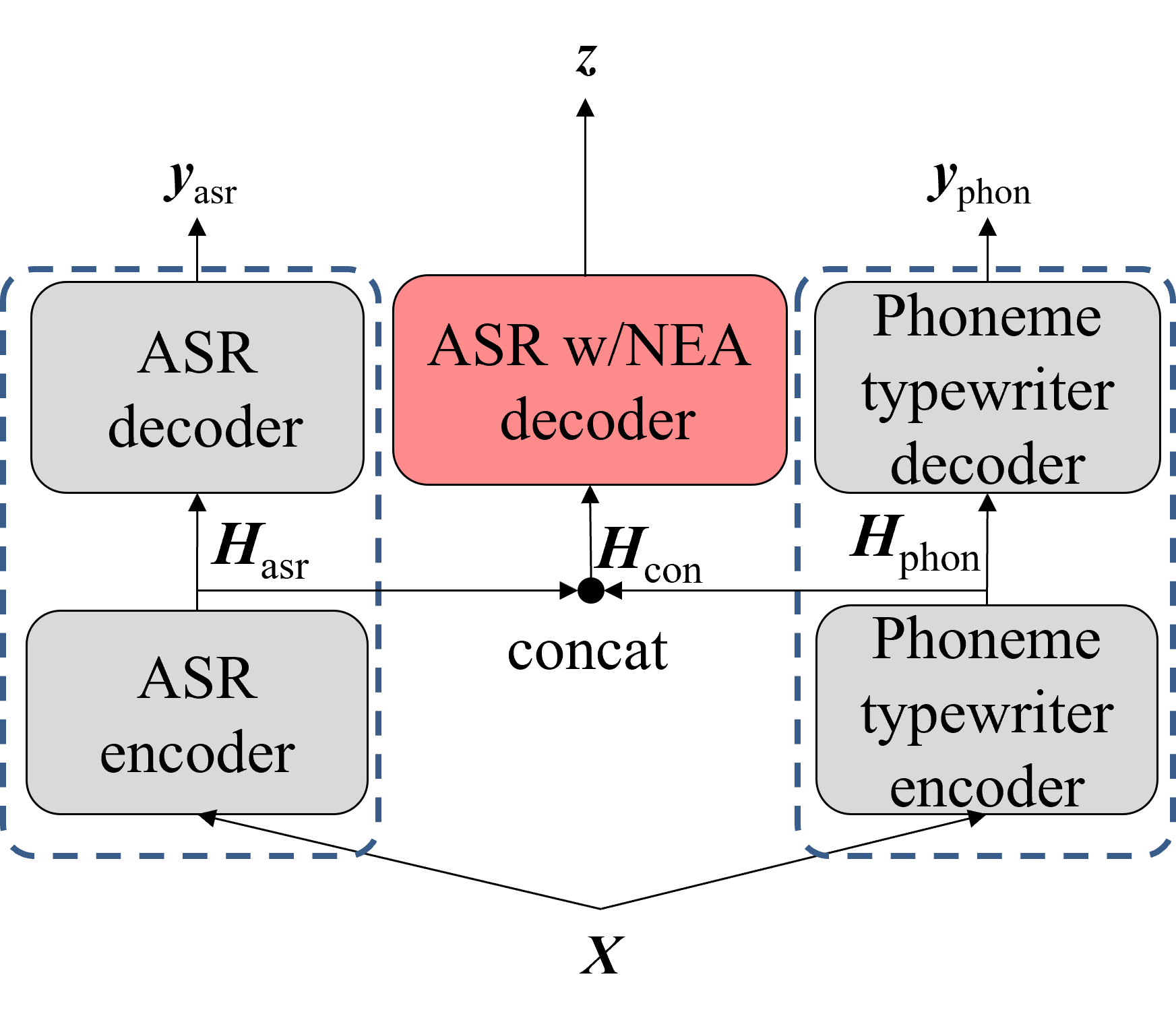}
         \vskip -0.1in
         \caption{Dual-encoder NEA.}
         \label{fig:dnea}
     \end{subfigure}
     \hfill
    \vskip -0.1in
    \caption{The network architectures of the NEA models.}
    \label{e2eclm}
    \vskip -0.1in
\end{figure}

\subsection{Single-encoder NEA}
\label{ssec:clm}
The S-NEA model with a single-encoder–decoder network is shown in Figure \ref{fig:snea}. This is the same as the one used in conventional ASRs, as described in Section~\ref{sec:Preliminary}. The only difference is that it estimates  \begin{math}\bm{z}\end{math} instead of $\bm{y}_S$ as follows:
\begin{equation}
\vspace*{-1mm}
\label{eq:clm}
P_{\text{}}(\bm{z} \mid \bm{X})=\prod_{s=1}^{|\bm{z}|} P\left(z_{s} \mid \bm{z}_{s-1}, \bm{H}\right).
\end{equation}
Note that since the phoneme sequences are added only to NE words using the special tokens SNE and ENE in the training data, the number of phoneme sequences in the training data may not be sufficient. This may reduce the robustness of NE words that contain rarely appearing phonemes in the training data.

\subsection{Dual-encoder NEA}
\label{ssec:sclapt}

The D-NEA consists of two pre-trained encoder-decoder sub-networks for a conventional ASR, and a phoneme typewriter (PT), and an NEA decoder sub-networks, as shown in Figure \ref{fig:dnea}. 
The PT is trained to estimate phoneme sequences instead of token sequences~\cite{chan16c_interspeech,Toshniwal2017MultitaskLW}.
By integrating these sub-networks, the D-NEA is expected to recognize a word even when the word includes phonemes that rarely appear during training. The two encoder-decoder sub-networks have the same network architecture as the S-NEA. 
Conventional ASR sub-networks estimate token sequences without phonemes, \begin{math}\bm{y}_{\mathrm{asr}}\end{math}, while the PT sub-network estimates phoneme sequences without tokens, \begin{math}\bm{y}_{\mathrm{p}}\end{math}.
The outputs of the decoder parts for the conventional ASR and  PT, \begin{math}\bm{H}_{\mathrm{asr}}\end{math} and \begin{math}\bm{H}_{\mathrm{p}}\end{math}, are concatenated into
\begin{math}
  \bm{H}_{\mathrm{con}} = \mathrm{concat}(\bm{H}_{\mathrm{asr}}, \bm{H}_{\mathrm{p}}).
\end{math}
The NEA decoder takes
\begin{math}\bm{H}_{\mathrm{con}}\end{math}
as an input, and estimates
\begin{math}\bm{z}\end{math}.
The model was trained in two stages. First, the ASR and PT sub-networks were trained, and then the NEA decoder was trained while freezing the trained ASR and PT sub-networks. 
Note that \begin{math}\bm{y}_{\mathrm{asr}}\end{math} and \begin{math}\bm{y}_{\mathrm{p}}\end{math} are used to pre-train the conventional ASR and the PT sub-networks, and they are discarded in the later processing, such as decoding.

\subsection{Phoneme similarity estimation and error correction}
\label{ssec:similarity}

Algorithm~\ref{similarity} shows the phoneme similarity estimation.
From the output of ASR with the NEA model, \begin{math}\bm{z}\end{math}, the $n$-th phoneme sequence, $\bm{z}_{\text{p}_n}$ is extracted based on the special tokens, SNE and ENE (line 2 in Algorithm~\ref{similarity}).
Then, phoneme similarity, $r$, between \begin{math}\bm{z}_{\text{p}_n}\end{math} and each entry in the dictionary, \begin{math}\bm{z}_{\text{d}_i} (i=1,\cdots,I)\end{math}, is calculated (lines 3-5 in Algorithm~\ref{similarity}). The phoneme similarity, $r$, is calculated using Gestalt pattern matching \cite{ratcliff1988pattern} described as follows:
\begin{equation}
\label{matching}
r =  \mathrm{Sim}(\bm{z}_{\text{d}_i}, \bm{z}_{\text{p}_n}) = \frac{2K}{|\bm{z}_{\text{d}_i}| + |\bm{z}_{\text{p}_n}|} \,\, (0 \le r \le 1),
\end{equation}
where \begin{math}K\end{math} denotes the number of the matched phonemes.
A value of $r = 1$ means that the two phoneme sequences are perfectly matched, while a value of $r = 0$ means that the two phoneme sequences are perfectly different.
Then, the NE with the highest phoneme similarity, $r_{\text{max}}$, and its index $\text{maxi}$, in the dictionary are selected (line 6 in Algorithm~\ref{similarity}).
Additionally, a threshold, $V_{th}$, is introduced to improve the robustness. If the phoneme similarity is greater than $V_{th}$, the corresponding NE, $y_{\text{NE}_n}$,  is replaced with that of the dictionary, $d_{\text{NE}_\text{maxi}}$, and the SNE, ENE, and phoneme sequences are deleted (line 8 in Algorithm~\ref{similarity}). Otherwise, the NE, $y_{\text{NE}_n}$, is not replaced, and the special token and phoneme sequence are deleted (line 10 in Algorithm~\ref{similarity}). This threshold prevents the proposed model from replacing an NE word with a non-NE word.

\begin{figure}[!t]
\vspace*{-3mm}
\begin{algorithm}[H]
\caption{Phoneme similarity estimation}
\label{similarity}
\begin{algorithmic}[1]
    \FOR {$n = 1,\cdots N$}
    \STATE {$\bm{z}_{\text{p}_n} = \text{ExtractPhoneme}(\bm{z},n)$}
    \FOR {$i = 1, \cdots, I$}
    \STATE {$r[i] = \text{Sim}(\bm{z}_{\text{d}_i}, \bm{z}_{\text{p}_n})$}
    \ENDFOR
    \STATE {[$r_{\text{max}}, \text{maxi}] = \text{max}(r)$}
    \IF {$r_{\text{max}} > V_{th}$}
        \STATE {$\bm{z} = \text{ReplaceNEword-and-DeletePhoneme}(\bm{z},\bm{d}_{\text{NE}_{\text{maxi}}},n$)}
    \ELSE {}
        \STATE {$\bm{z} = \text{DeletePhoneme}(\bm{z},n)$}
    \ENDIF
    \ENDFOR
    \STATE {$\bm{\textbf{return}} \, \bm{z}$ }
\end{algorithmic}
\end{algorithm}
\vspace*{-8mm}
\end{figure}

\section{Experiments}

We evaluated the proposed S-NEA and D-NEA models in terms of performance, scalability, and robustness of the proposed method. 
We then analyzed the recognition results for personal names to confirm the effectiveness of the proposed method for enharmonic words. Furthermore, the number of NE words, $I$, registered in the dictionary and the effect of the threshold, $V_{th}$, on performance were examined to verify the scalability and robustness of the proposed method.

\vspace*{-1mm}
\subsection{Experimental setup}

The input feature was an 80-dimensional Mel-scale filter-bank feature with a window size of 512 samples and a hop length of 128 samples. The sampling frequency was 16 kHz. SpecAugment \cite{specaug} was then used.
All encoders in the S-NEA and D-NEA models had the same network structure. Each encoder comprised two convolutional layers with stride two, a 512-dimensional linear projection layer, and a positional encoding layer, followed by 12 Conformer layers with 2,048 linear units and layer normalization. The decoder had six transformer layers with 2,048 units. Both the decoders in the S-NEA and D-NEA models had the same network structures. 
The dimension size of the attention was 512 with 8-multihead attentions.
The proposed model was trained on 50 epochs using the Adam optimizer at a learning rate of 0.0015 with 15,000 warm-up steps.


We used a customized dataset for training. The training dataset consisted of the Corpus of Spontaneous Japanese~\cite{csj}, a multiple speaker speech database developed by the Advanced Telecommunications Research Institute International (ATR-APP)~\cite{KUREMATSU1990357}, and our in-house dataset. Since the CSJ does not provide "person name" tags, we used MeCab~\cite{kudo-etal-2004-applying}, a morphological analysis tool, to tag the training data. Our in-house dataset consists of 93 hours of single-speaker speech data collected from three different locations, including multiple scenarios, such as meetings and morning assemblies.\footnote{Our in-house dataset is not released for confidentiality and privacy reasons.}
We used two evaluation datasets, our in-house evaluation dataset and CSJ eval3. Personal names were present in 5.8\% of the total 4,838 utterances in our in-house evaluation dataset. CSJ eval1 and eval2 were excluded because they contained too few personal names. The character error rate (CER) for the token sequences of the target NEs (CER-NE) and all character sequences (CER-all) were calculated to evaluate the effect of the proposed method. The CER-NE was calculated within a subset of the target NEs, defined as follows:
\begin{equation}
\text{CER-NE} = \frac{S_\text{NE} + I_\text{NE} + D_\text{NE}}{N_\text{NE}},
\end{equation} 
where $S_\text{NE}$, $I_\text{NE}$, $D_\text{NE}$, and $N_\text{NE}$ denote the number of substitutions, insertions, deletions, and the total number of tokens in person names, respectively.
The threshold, $V_{th}$, described in Section~\ref{ssec:similarity} was set to 0.8. 
We used the ESPnet~\cite{espnet} toolkit. 

\subsection{Comparison between S-NEA and D-NEA}

The result of the comparison between the S-NEA and D-NEA models is shown in Table \ref{mainresults}. For the two evaluation sets, both of the proposed models outperformed the baseline. In particular, the CER-NE for the D-NEA was 35.7\% lower than the baseline. Furthermore, the CER-all was also decreased due to the improved CER-NE. The CER-all for the S-NEA in the CSJ eval3 was 3.4\%, which outperformed the state of the art~\cite{Karita2021ACS}. Comparing the two proposed methods, the D-NEA outperformed the S-NEA for the CER-NE for both evaluation datasets. Although the CER-all of the D-NEA were slightly worse than those of the S-NEA, the improved CER-NE is expected to enhance usability because personal names are extremely important for specific communities and tasks, as described in the introduction.

\begin{table}[t]
\caption{Main results comparing the CER-all and CER-NE for the in-house dataset and CSJ eval3.}
\vspace*{-6mm}
\label{mainresults}
\begin{center}
\begin{tabular}{@{}c|cc|cc|cc}
\hline
    & \multicolumn{2}{c|}{In-house} & \multicolumn{2}{c|}{CSJ eval3} & \multicolumn{2}{c}{Average}\\
Method & All & NE & All & NE & All & NE \\
\hline
Baseline & 6.9 & 33.5 & 3.6 & 38.7 & 5.4 & 35.0\\
S-NEA  & \textbf{6.8} & 23.9 & \textbf{3.4} & 33.6 & \textbf{5.1} & 26.6\\
D-NEA  & 7.0 & \textbf{22.2} & 3.5 & \textbf{23.4} & 5.3 & \textbf{22.5}\\
\hline
\end{tabular}
\end{center}
\vspace*{-4mm}
\end{table}

\begin{figure}[t]
\begin{center}
{
\resizebox*{8.0cm}{!}{\includegraphics{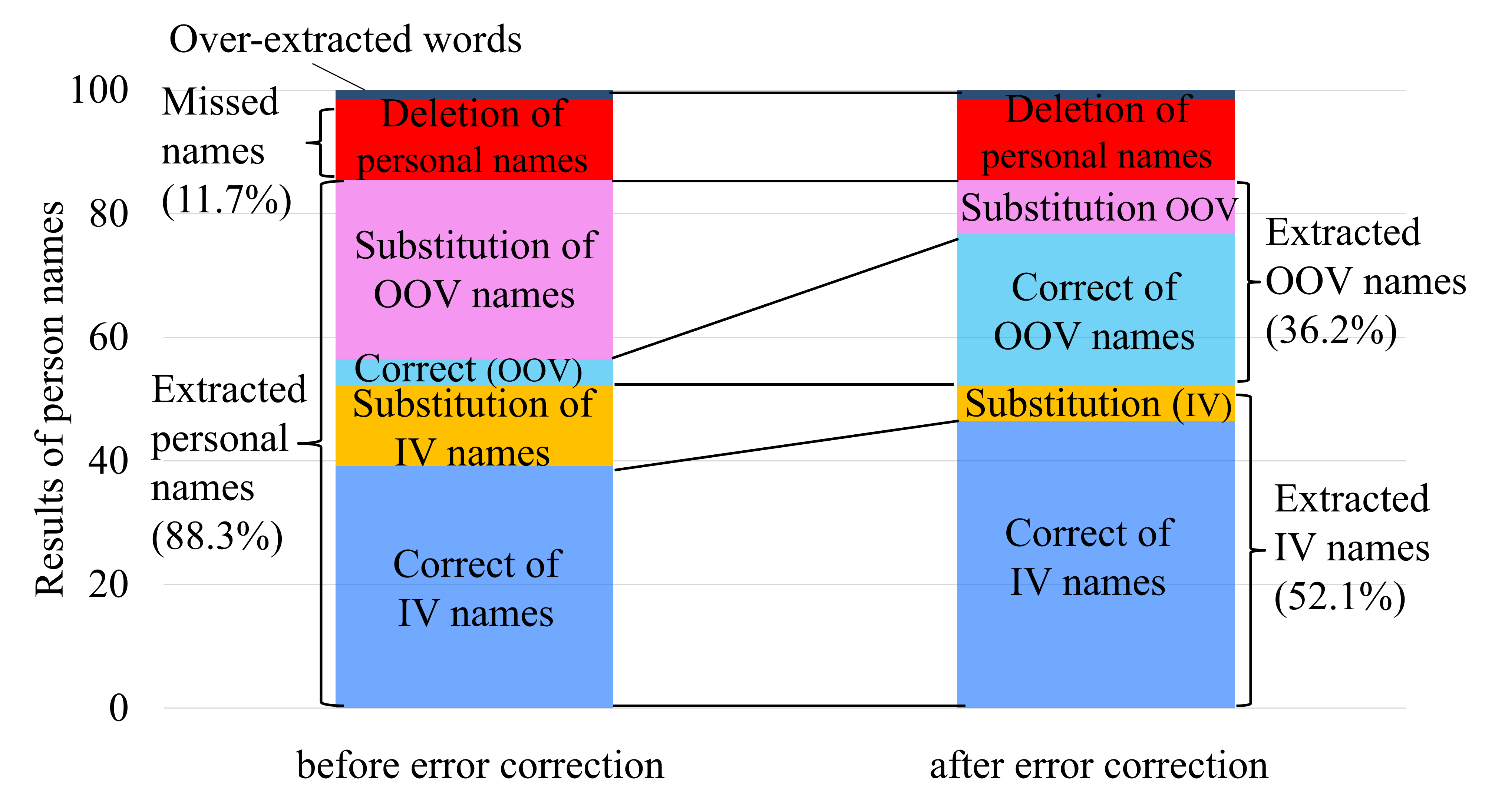}}}
\vspace*{-8mm}
\caption{Breakdown of the results of the target named entities.}
\label{analysis} 
\end{center}
\vspace*{-5mm}
\end{figure}

We analyzed the effectiveness of the proposed method in detail. Figure~\ref{analysis} shows the results for the in-house evaluation dataset using the D-NEA model before and after the error correction. 88.3\% of the personal names in the evaluation set were correctly extracted. Of these, 52.1\% were IV personal names included in the training data, and the remaining 36.2\% were OOV personal names. 
The substitution errors of the IV personal names were 13.0\% before the error correction, which was caused by the enharmonic word problem, and it reduced to be half after the correction using the dictionary, as shown in orange in Figure~\ref{analysis}.
Most OOV personal names resulted in substitution errors before the error correction, whereas they were well recognized after the correction with the dictionary, as shown in pink in Figure~\ref{analysis}. 

\subsection{Influence of the dictionary size}

\begin{figure}[t]
\begin{center}
{
\resizebox*{6.0cm}{!}{\includegraphics{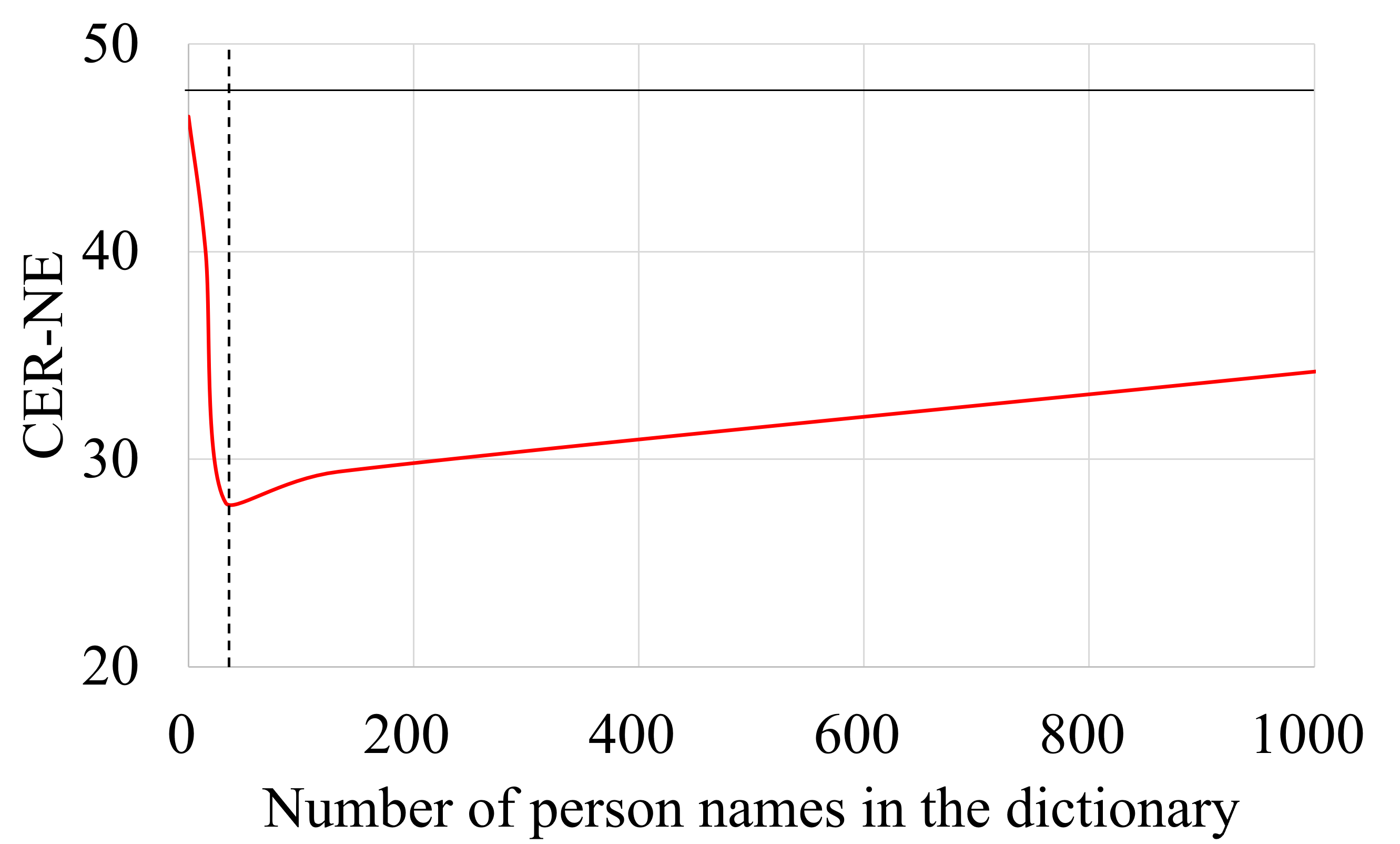}}}
\vspace*{-3mm}
\caption{Effect of the number of words in the dictionary.}
\label{numwords} 
\end{center}
\vspace*{-4mm}
\end{figure}

We tested the effect of the number of personal names, $I$, as described in Eq.~\eqref{eq:dictionary} in the dictionary for the CER-NE, since $I$ tends to be large unless the user intentionally erases names that have been registered in the dictionary in the past.
Figure~\ref{numwords} shows the impact of $I$ on the CER-NE for the in-house evaluation dataset using the D-NEA model.
Since the proposed method is intended to be used in a specific community, a subset of one location was used.
When the dictionary was not used ($I$ = 0), the CER-NE was 46.5\%. 
As the number of personal names in the dictionary increased, the CER-NE improved, because the number of enharmonic personal names (IV and OOV) decreased.
The CER-NE was the lowest when the dictionary contained the exact same personal names ($I$ = 33) as those used in the evaluation dataset because the number of enharmonic personal names was zero.
As $I$ became even larger, the CER-NE gradually increased, but the CER-NE was still better than the case without the dictionary, even when registering 1,000 personal names ($I$ = 1,000) in the dictionary. This shows the robustness of the proposed method against the increased number of registered personal names.
The analysis of the data obtained from the users of the proposed system showed that only 5.7\% of the utterances contained the personal names. Therefore, the computational cost did not change for the remaining 94.3\% of the utterances. The additional computational cost for the utterances containing personal names 
was only 0.4\% of the baseline for a dictionary of 1,000 personal names.

\subsection{Effect of the threshold}
\label{sec:th}

The effect of the phoneme similarity threshold, $V_{th}$, described in Section~\ref{ssec:similarity}, was examined. 
Table~\ref{thresh} shows the effect of the phoneme similarity threshold, $V_{th}$, on the CER-NE for the in-house evaluation dataset using D-NEA.
When $V_{th}$ = 0.5, the CER-NE was the smallest.
When $V_{th}$ = 0.0, all extracted personal names were replaced with the personal names from the dictionary, which caused an error when the NEA model mis-extracted a non-personal name as a personal name. Conversely, when $V_{th}$ = 1.0, personal names are replaced only when the extracted phoneme sequences exactly matched the phoneme sequence in the dictionary, so even a small error in the phoneme sequence output by the NEA model resulted in the degradation of the CER-NE.
Therefore, the threshold, $V_{th}$ should be set by considering the balance between these two errors.

\begin{table}[t]
\caption{Effect of the threshold.}
\vspace*{-7mm}
\label{thresh}
\begin{center}
\begin{tabular}{@{}c|cc}
\hline
Threshold $V_{th}$ & CER-all & CER-NE \\
\hline\hline
0.0 & \textbf{7.0} & 27.6\\
0.5 & \textbf{7.0} & \textbf{19.8}\\
0.8 & \textbf{7.0} & 22.2\\
1.0 & 7.1 & 28.6\\
\hline
\end{tabular}
\end{center}
\vspace*{-4mm}
\end{table}

\section{Conclusion}

This paper presented a retraining-free customizable E2E-ASR model consisting of ASR with an NEA model trained on a corpus tagged with target NEs, a dictionary for a target NE containing enharmonic words, phoneme similarity estimations, and error corrections. The proposed method successfully improved recognition of both IV and OOV enharmonic personal names. We plan to extend the proposed method so that it can be customized for not only personal names but also other NEs.

\clearpage

\bibliographystyle{IEEEtran}
\bibliography{mybib}
\end{document}